\documentclass{jors}

\definecolor{mygray}{gray}{0.6}

\usepackage{amsmath,amssymb,amsfonts}
\usepackage{mathtools}
\usepackage{bm}
\usepackage[symbol]{footmisc}
\usepackage[citestyle=numeric,bibstyle=authoryear,backend=bibtex]{biblatex}
\begin{document}

%{\bf Software paper for submission to the Journal of Open Research Software} \\

%To complete this template, please replace the blue text with your own. The paper has three main sections: (1) Overview; (2) Availability; (3) Reuse potential. \\

%Please submit the completed paper to: editor.jors@ubiquitypress.com

%\rule{\textwidth}{1pt}

\section*{(1) Overview}

\vspace{0.5cm}

\section*{Title}
%\textcolor{blue}{The title of the software paper should focus on the software, e.g. “Text mining software from the X project”. If the software is closely linked to a specific research paper, then “Software from Paper Title” is appropriate. The title should be factual, relating to the functionality of the software and the area it relates to rather than making claims about the software, e.g. “Easy-to-use”.}
{\LARGE \textbf{Palabos-npFEM: Software for the Simulation of Cellular Blood Flow (Digital Blood)}}

\section*{Paper Authors}
%\textcolor{blue}{1. Last name, first name; (Lead/corresponding author first) \\
%2. Last name, first name; etc.}
1. Kotsalos, Christos (Lead/corresponding author)

2. Latt, Jonas

3. Chopard, Bastien

\section*{Paper Author Roles and Affiliations}
%\textcolor{blue}{1. First author role and affiliation \\
%2. Second author role and affiliation etc.}
1. C.K. performed the research, developed the majority of the computational framework, carried out the simulations and wrote the paper. Computer Science Department, University of Geneva.

Email: christos.kotsalos@unige.ch

2. J.L. wrote part of the computational framework, supervised the research and revised the manuscript. Computer Science Department, University of Geneva. 

Email: jonas.latt@unige.ch

3. B.C. conceived and supervised the research and revised the manuscript. Computer Science Department, University of Geneva.

Email: bastien.chopard@unige.ch

\section*{Abstract}
%\textcolor{blue}{A short (ca. 100 word) summary of the software being described: what problem the software addresses, how it was implemented and architected, where it is stored, and its reuse potential.}
Palabos-npFEM is a computational framework for the simulation of blood flow with fully resolved constituents. The software resolves the trajectories and deformed state of blood cells, such as red blood cells and platelets, and the complex interaction between them. The tool combines the lattice Boltzmann solver Palabos for the simulation of blood plasma (fluid phase), a finite element method (FEM) solver for the resolution of blood cells (solid phase), and an immersed boundary method (IBM) for the coupling of the two phases. Palabos-npFEM provides, on top of a CPU-only version, the option to simulate the deformable bodies on GPUs, thus the code is tailored for the fastest supercomputers. The software is integrated in the Palabos core library, and is available on the Git repository \url{https://gitlab.com/unigespc/palabos}. It offers the possibility to simulate various setups, e.g. several geometries and blood parameters, and due to its modular design, it allows external solvers to readily replace the provided ones.

\section*{Keywords}
%\textcolor{blue}{keyword 1; keyword 2; etc. \\
%Keywords should make it easy to identify who and what the software will be useful for.}
Palabos-npFEM; cellular blood flow simulations; digital blood; Palabos; npFEM; GPUs;

\clearpage

\setlength{\parindent}{15pt}
\section*{Introduction}
%\textcolor{blue}{An overview of the software, how it was produced, and the research for which it has been used, including references to relevant research articles. A short comparison with software which implements similar functionality should be included in this section.}
\noindent Palabos-npFEM is a highly versatile computational tool for the simulation of cellular blood flow (at the micrometre scale), focusing on high performance computing (HPC) without compromising accuracy or complexity.

Blood plays a vital role in living organisms, transporting oxygen, nutrients, waste products, and various kinds of cells, to tissues and organs. Human blood is a complex suspension of red blood cells (RBCs), platelets (PLTs), and white blood cells, submerged in a Newtonian fluid, the plasma. At physiological hematocrit (RBCs volume fraction), i.e. 35-45\%, in just a blood drop (about a $mm^3$) there are a few million RBCs, a few hundred thousand PLTs, and a few thousand white blood cells. An adult person has on average five litres of blood, and the cardiovascular system spans a length of $100,000~km$, 80\% of which consists of the capillaries (smallest blood vessels). Additionally, our blood vessels are characterised by a variety of scales, i.e. the diameter of arteries/veins ranges from few millimetres to few centimetres, the diameter of arterioles/venules ranges from few micrometres to few hundred micrometres, and the capillaries are about the size of a RBC diameter (about eight micrometres). It is obvious that a simulation at the micrometre scale of such a system (even a tiny part of it, e.g. an arteriole segment) is a multi-physics/multi-scales problem, with an extremely high computational cost.

Despite remarkable advances in experimental in-vivo and in-vitro techniques \cite{Tomaiuolo2014BiomechanicalMicrofluidics}, the type and detail of information provided remains limited. For example, while it is possible to track the collective behaviour of blood cells, it is up to now impossible to track individual trajectories and study the underlying transport mechanisms. Given the cellular nature of blood, its rheology can be deciphered by understanding how the various cells are moving and interacting with each other in both healthy and non-healthy humans. Numerical tools have significantly assisted in in-depth investigations of blood rheology, as they offer a controlled environment for testing a large number of parameters and classifying their effect \cite{Boudjeltia2020,Kotsalos_FatTails}. Furthermore, the amount of detail coming from the numerical simulations at the microscopic level (following individual cells) is unparalleled compared to the in-vitro/in-vivo counterparts.

The multi-physics nature of blood can be numerically described by decomposing this complex suspension into the fluid and solid phases.

Regarding the fluid phase, we are interested on solving the Navier-Stokes equations for the description of blood plasma. Thus the spatial velocity $\bm{v}(\bm{x},t)$ and pressure $p(\bm{x},t)$ fields in a body of Newtonian isothermal and incompressible fluid with reference configuration $B$ must satisfy the following equations for all $\bm{x} \in B_t$ and $t \geq 0$ \cite{gonzalez_stuart_2008}
\begin{equation} \label{eq:fluid_eqns}
    \begin{aligned}
        \rho_0 \left [ \frac{\partial}{\partial t}\bm{v} + \left ( \nabla^x\bm{v} \right )\bm{v} \right ] = \mu \Delta^x\bm{v} - \nabla^x p + \rho_0 \bm{b} \\
        \nabla^x \cdot \bm{v} = 0
    \end{aligned}
\end{equation}
where $\nabla^x, \Delta^x$ refer to the gradient and Laplacian with respect to the spatial coordinates $\bm{x}$ for any fixed time $t \geq 0$, $\rho_0$ is the fluid density, $\mu$ is the fluid dynamic viscosity, and $\bm{b}$ is a prescribed spatial body force field per unit mass (such as gravity, an immersed surface, etc.).

Regarding the solid phase, for the resolution of the deformable blood cells and their trajectories, we are solving the Elastodynamics equation \cite{gonzalez_stuart_2008}, a non-linear equation, referring to any kind of solid body, i.e. membrane or full 3D solid. By convention, we call $B$ the reference (undeformed) configuration, and $B'$ the deformed configuration. The points $\mathbf{X} \in B$ are called material coordinates, and the points $\bm{x} \in B'$ are called spatial coordinates. The deformation of a body from a configuration B onto another configuration $B'$ is described by a function $\bm{\phi} : B \rightarrow B'$, which maps each point $\mathbf{X} \in B$ to a point $\bm{x} = \bm{\phi}(\mathbf{X}) \in B'$. We call $\bm{\phi}$ the deformation map. The motion $\bm{\phi}(\mathbf{X},t)$ of an elastic body must satisfy the following equation for all $\mathbf{X} \in B$ and $t \geq 0$ \cite{gonzalez_stuart_2008}
\begin{equation}  \label{eq:solid_eqns}
    \rho_0 \bm{\ddot{\phi}} = \nabla^X \cdot \mathbf{P} + \rho_0 \bm{b}_m 
\end{equation}
where $\nabla^X$ refers to derivatives of material fields with respect to the material coordinates $X_i$ for any fixed $t \geq 0$, $\rho_0(\mathbf{X})$ denotes the mass density of the elastic body in its reference configuration, $\mathbf{P}(\mathbf{X},t)$ is the first Piola-Kirchhoff stress field, and $\bm{b}_m(\mathbf{X},t)$ is the material description of the spatial body force field $\bm{b}(\bm{x},t)$. Equation (\ref{eq:solid_eqns}) is essentially the conservation of linear momentum (Newton's $2^{nd}$ law of motion), while the balance of angular momentum is automatically satisfied from the symmetry of the Cauchy stress field (by definition of elastic bodies). With minor modifications of equation (\ref{eq:solid_eqns}), one can readily simulate a viscoelastic material, i.e. a body that exhibits both viscous and elastic characteristics.

For more details on the continuum equations for both fluids and solids, the reader should consult the seminal work by Gonzalez and Stuart \cite{gonzalez_stuart_2008}.

The two phases are aware of each other, simply by the last term in equations (\ref{eq:fluid_eqns}) \& (\ref{eq:solid_eqns}), i.e. through the body force field $\bm{b}$. This leads to a very efficient Fluid-Structure/Solid Interaction (FSI), since there is no need for the fluid/solid meshes to conform, allowing for different discretisation strategies per phase.

\begin{figure}[h]
    \centering
    \includegraphics[width=0.8\textwidth]{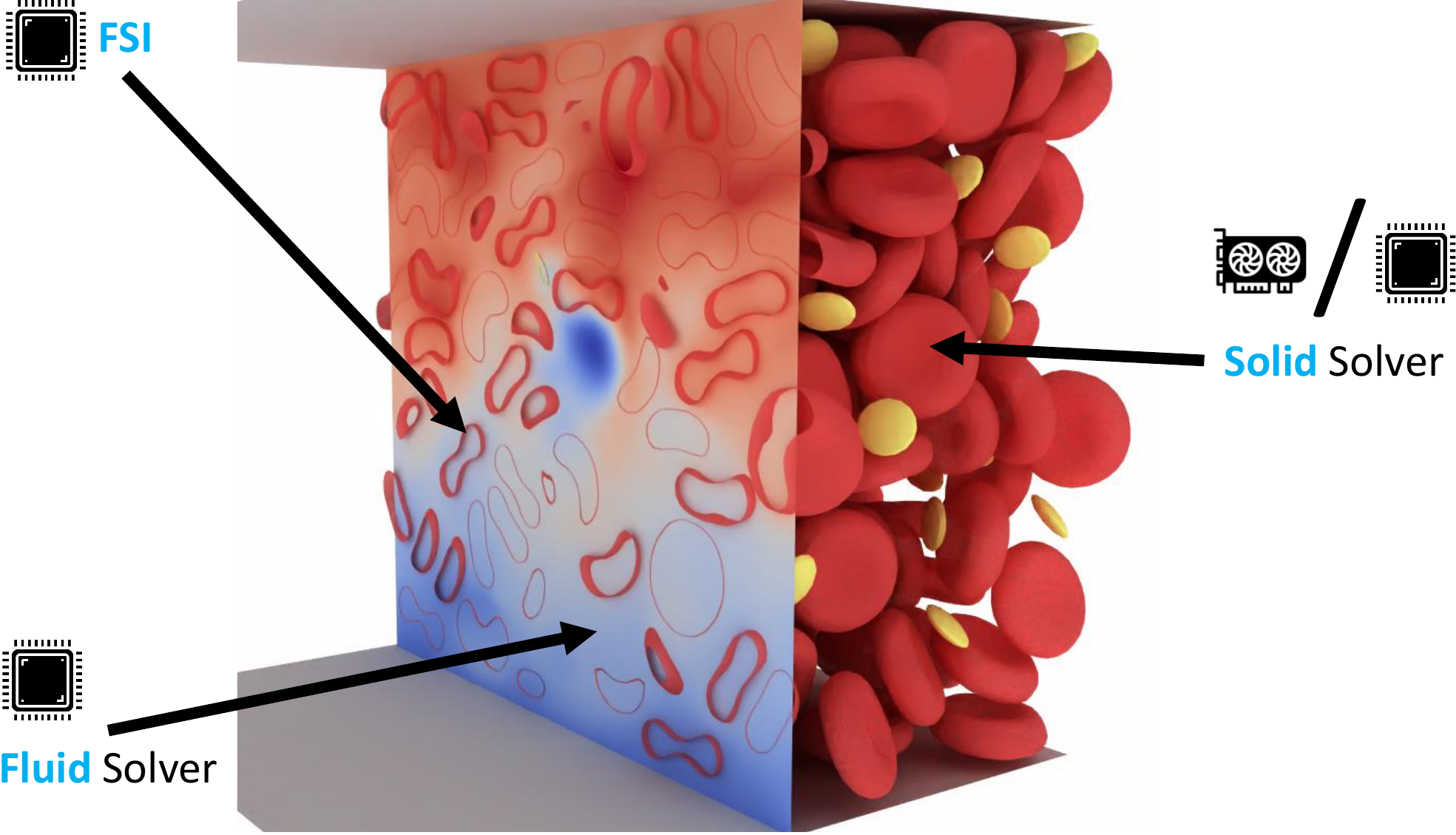}
    \caption{Digital Blood simulated in Palabos-npFEM (a simulation snapshot). The red bodies are the RBCs, while the yellow bodies represent the PLTs. One can observe the different phases composing this complex suspension. We provide the possibility to run the solid solver either on CPUs or GPUs (as depicted by the GPU/CPU icons), while the rest of the solvers are CPU-only versions.}
    \label{fig:solvers_cross_section}
\end{figure}

Palabos-npFEM solves equations (\ref{eq:fluid_eqns}) \& (\ref{eq:solid_eqns}) in a modular way, and performs the FSI through the Immersed Boundary Method (IBM). By modularity, we mean the complete decoupling of the solvers, i.e. the resolution of fluid phase is ``unaware'' of the resolution of the solid phase and vice-versa. Our software framework takes care of communication whenever needed in relationship with the FSI. Figure \ref{fig:solvers_cross_section} presents a snapshot from a Palabos-npFEM simulation, and shows the different solvers involved in the numerical representation of blood. The Navier-Stokes equations (\ref{eq:fluid_eqns}) are solved indirectly through the lattice Boltzmann method (LBM) as implemented in Palabos\footnote[1]{\url{https://palabos.unige.ch}} \cite{PalabosArticle}. Palabos stands for Parallel Lattice Boltzmann Solver. It is an open source software maintained by the Scientific and Parallel Computing Group (SPC) at the University of Geneva. The elastodynamics equation (\ref{eq:solid_eqns}) is solved by the nodal projective finite elements method (npFEM) \cite{Kotsalos2019}. The npFEM is a mass-lumped linear FE solver that resolves both the trajectories and deformations of blood cells with high accuracy. The solver has the capability of capturing the rich and non-linear viscoelastic behaviour of any type of blood cells as shown and validated in Kotsalos et al. \cite{Kotsalos2019}. The IBM \cite{Peskin1972FlowMethod,Ota2012LiftSimulations} for the coupling of the solid \& fluid phases, i.e. the computation of term $\bm{b}$ in equations (\ref{eq:fluid_eqns}) \& (\ref{eq:solid_eqns}), is implemented in the Palabos library. The modular system allows different spatial discretisations, i.e. the fluid domain is discretised into a regular grid with spacing $\Delta x$ in all directions, and the solid bodies are discretised into triangular surface meshes. The data exchange between solvers of different discretisation is handled through interpolation kernels, as dictated by the immersed boundary method. Moreover, the temporal discretisation is solver dependent. Given the explicit nature of LBM and the implicit/semi-implicit nature of npFEM, the latter can in principle handle larger time steps.

Our modular design contrasts with monolithic approaches, which solve both phases through one system of discretised equations, and thus use one solver. Monolithic approaches can be considered as well computational frameworks whose design entangles the various solvers for performance purposes. The former approach, i.e. a single solver to deal with every phase, includes mainly tools that use dissipative particle dynamics \cite{Rossinelli_2015}. The main advantage of a monolithic design is the performance gain and a more straightforward FSI (in terms of coupling difficulty). However a single solver potentially falls short of satisfactorily addressing all the physics present in a complex phenomenon. In the modular approach, there is the freedom to choose well optimised solvers to address the different phases, which leads to higher fidelity models. Of course, the coupling of completely independent solver streams can introduce performance penalties and possibly an over-complicated code. However, we have shown \cite{Kotsalos2020}, that our modular design results in a minimal performance loss.

The development of such a solver requires a multi-physics and multi-scale approach as it involves fluid and solid components that may be optimised through a description at different temporal and spatial scales. To ensure flexibility and efficiency, code performance and re-usability are central issues of our modular tool. According to the principles proposed in MMSF (Multi-scale Modelling and Simulation Framework) \cite{chopard-PhilA:2013, borgdorff:jpdc13, Borgdorff-PhilA:2013}, our cellular blood flow computational tool is built on a fluid solver, and a deformable solid bodies solver, whose implementation is potentially left to the preference of the scientists. Here, however, we propose a specific choice. The coupling of the two solvers is realised through a separate interface, which handles all the needed communication. Note that this coupling interface acts independently of the details of the fluid and solid solvers. It only requires data representing physical quantities which are computed by the two solvers, thus ensuring the independence with respect to the chosen numerical methods \cite{Kotsalos2020}.

For a detailed presentation of the numerical methods used in Palabos-npFEM and its HPC-centric design, the reader should consult Kotsalos et al. \cite{Kotsalos2019,Kotsalos2020}. Here we focus on the software issues.

\section*{Implementation and architecture}
%\textcolor{blue}{How the software was implemented, with details of the architecture where relevant. Use of relevant diagrams is appropriate. Please also describe any variants and associated implementation differences.}
\noindent Palabos-npFEM is build on top of two independent solvers, i.e. fluid and solid solvers, and couples them through the FSI module (see Figure \ref{fig:solvers_cross_section}). The choice of the particular solvers, namely Palabos and npFEM, is crucial for high fidelity and performant simulations. However, other users of the code could extend it by replacing the particular choices by alternatives, e.g. opting for a mass-spring-system solid solver instead of a FEM one. The alternative solvers need to be similarly parallelisable through domain decomposition and allow interaction with solid particles through an immersed boundary method.

Palabos \cite{PalabosArticle} is an open-source library for general-purpose computational fluid dynamics based on the lattice Boltzmann method. Palabos is written in C\texttt{++} with extensive use of the Message Passing Interface (MPI). MPI is the library that handles parallelisation across multiple cores and nodes of a supercomputer/cluster. Palabos supports CPU-only hardware.

The npFEM solver \cite{Kotsalos2019} is an open-source finite element solver written in C\texttt{++} with support for openMP (for multi-core machines) and CUDA (GPU-support). CUDA is a general purpose parallel computing platform and programming model for NVIDIA GPUs. Thus npFEM provides two actively supported branches, i.e. CPU and GPU versions (as summarised in Figure \ref{fig:solvers_cross_section}). The npFEM solver is derived from a heavily modified version of the open-source library ShapeOp\footnote[3]{\url{https://www.shapeop.org/}} \cite{Bouaziz2014ProjectiveSimulation}. The different naming originates from the fact that the modifications make the solver an FEM solver instead of a computer graphics tool, as ShapeOp is initially intended for. In more details:
\begin{itemize}
    \item We have changed radically the original kernel for advancing the bodies in time. Our approach follows the redesigned projective dynamics approach as described in Liu et al. \cite{Liu2017Quasi-NewtonMaterials}.
    \item In Computer Graphics, the solvers are approximating Newton's equations (conservation of linear and angular momenta), reducing the computational cost. Our solver is not approximating Newton's equations, but we provide a converged solution, focusing on accuracy and physically correct states.
\end{itemize}
For legacy reasons, we have decided to keep the ShapeOp code structure and file naming. This approach allows the users to better understand our extensions, perform their own, and most importantly to benefit from both communities, namely the computational science and computer graphics ones.

\begin{figure}[h]
    \centering
    \includegraphics[width=\textwidth]{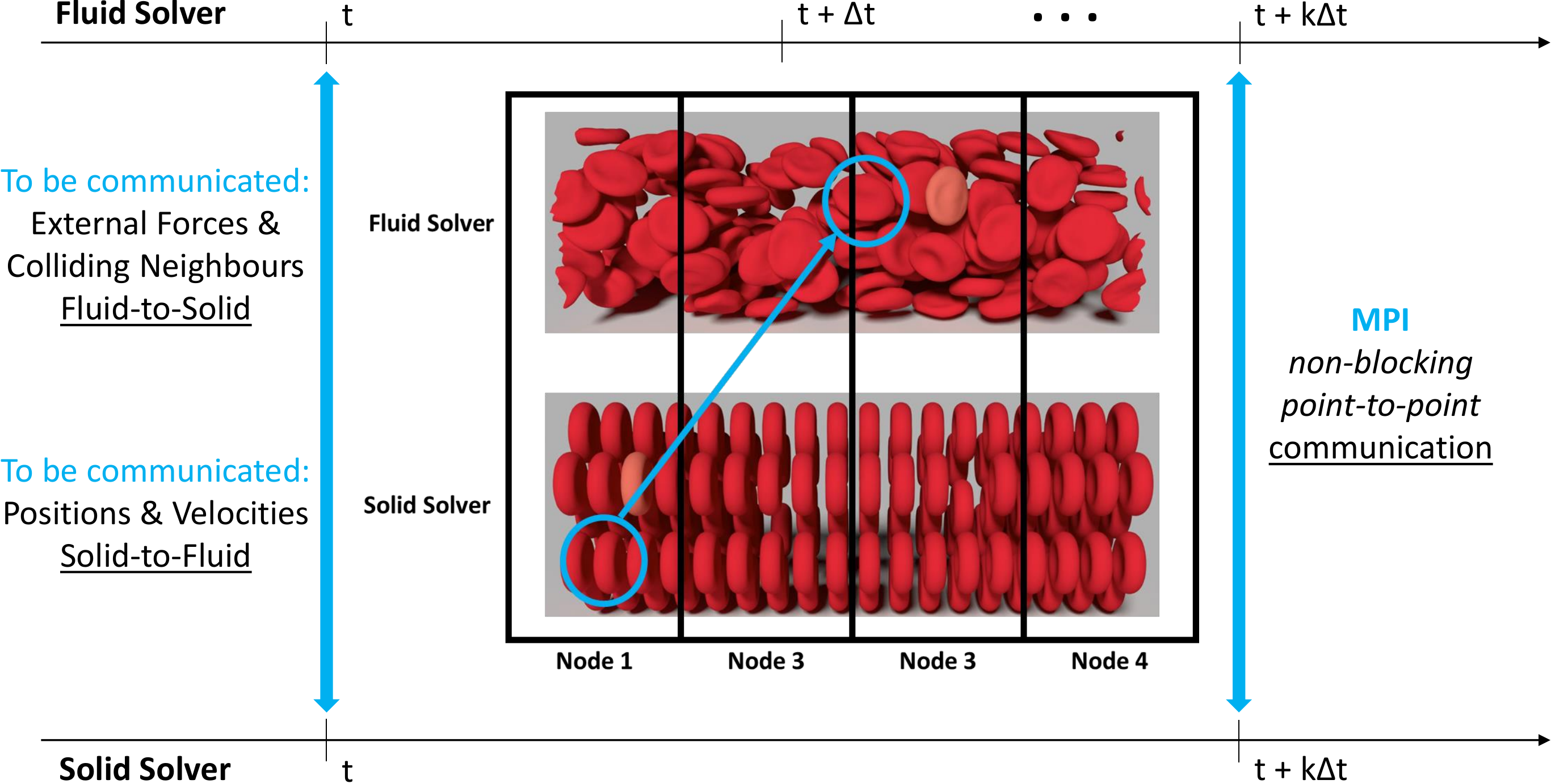}
    \caption{Modular design of Palabos-npFEM, where the different solvers are independent execution streams (see arrows through time). The coupling demands data exchange (two-way arrows) which is performed mainly through MPI communication. The inset shows a simulation snapshot and the static allocation of the different domains in the available infrastructure. Data that reside in different memory domains must be communicated through MPI.}
    \label{fig:solvers_independence}
\end{figure}

\begin{figure}[h]
    \centering
    \includegraphics[width=0.55\textwidth]{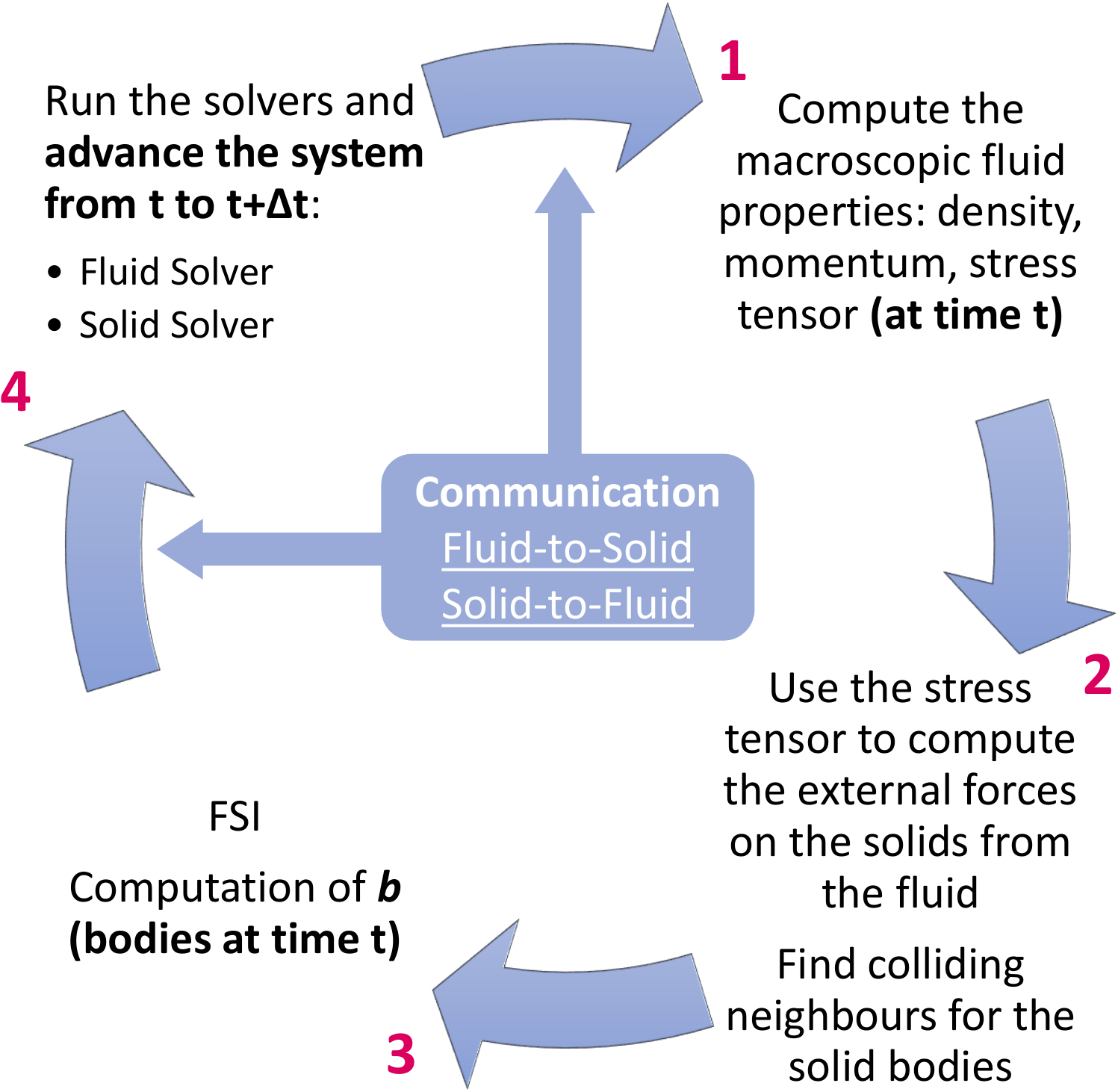}
    \caption{Execution order for various operations in Palabos-npFEM. The execution order is dictated by the fluid-solid interaction (FSI), i.e. the computation of term $\bm{b}$ in equations (\ref{eq:fluid_eqns}) \& (\ref{eq:solid_eqns}). In case of different temporal discretisation, some steps are automatically deactivated.}
    \label{fig:solvers_steps}
\end{figure}

\begin{figure}[h]
    \centering
    \includegraphics[width=0.8\textwidth]{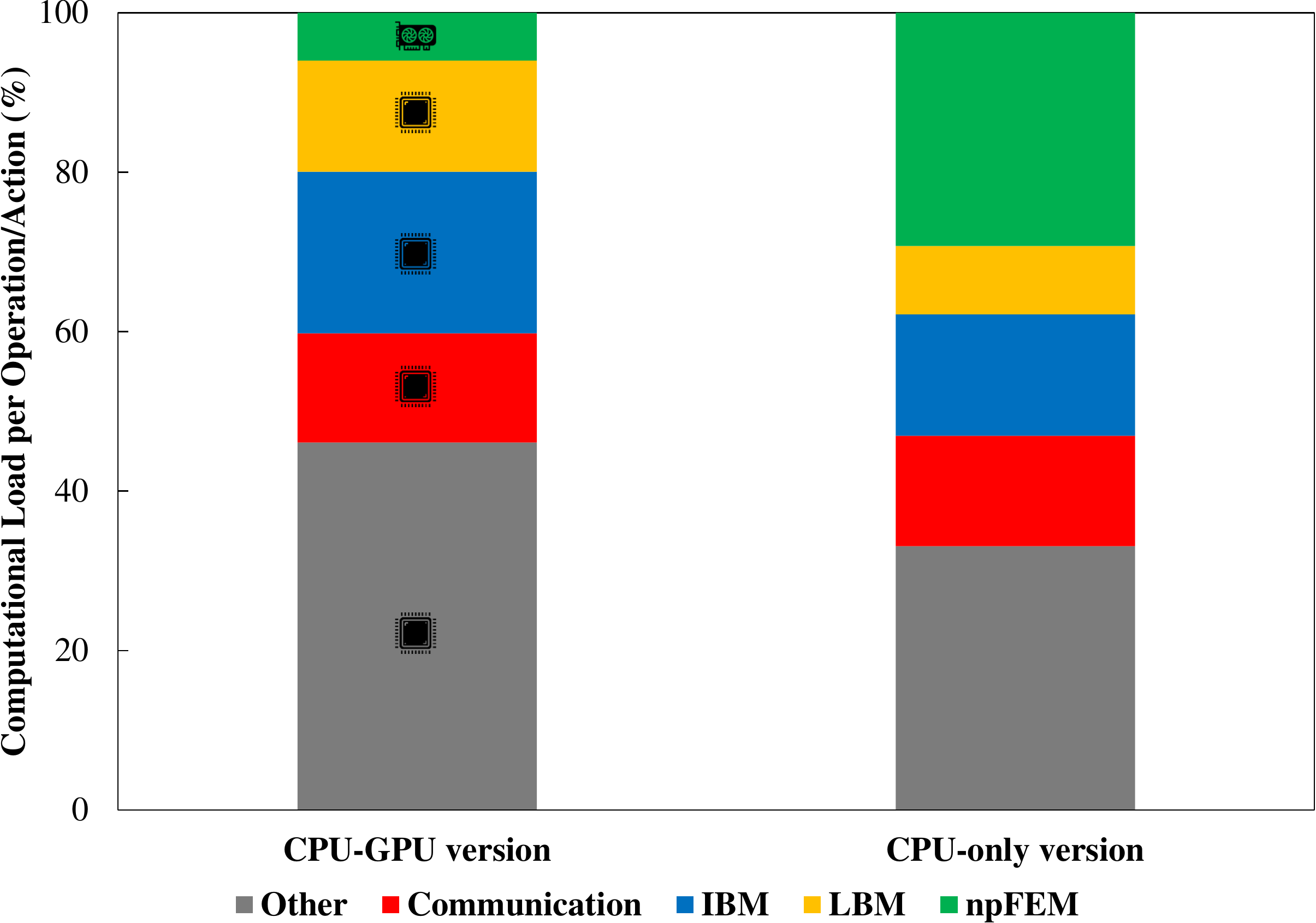}
    \caption{Average computational load per operation/action (\%) in both CPU-only and CPU-GPU versions. The ``Other'' category gathers the collision detection, the computation of forces on solids, and various book-keeping operations. The graph refers to cases at 35\% hematocrit.}
    \label{fig:stacked_percentages}
\end{figure}

Figure \ref{fig:solvers_independence} presents the realisation of this modular design of Palabos-npFEM. Indeed, the solvers possess independent execution streams (their implementation details are black-boxes to the end user). Their coupling consists of exchanging data, i.e. external forces \& colliding neighbours from fluid-to-solid, and positions \& velocities from solid-to-fluid. Our framework takes care of this exchange and quietly handles the parallelisation through MPI. By parallelisation, we mean the load balancing and the allocation of the solvers to the available hardware. The load balancing follows a straightforward and static (not changing with time) allocation to the available hardware. Regarding the fluid solver, the domain is decomposed into multiple non-overlapping sub-domains, where each one is handled by a single CPU-core. Regarding the solid solver, the blood cells are distributed to the available CPU-cores or GPUs, depending on the npFEM version that the user chooses. Additionally, every blood cell fits entirely in one hardware unit, i.e. either a CPU-core or a GPU-CUDA-block \cite{Kotsalos2020}, but every hardware unit may receive more than one blood cell. Both solvers exploit at the same time all the available hardware, i.e. there is no infrastructure grouping for the fluid or solid solvers. The data exchange is performed through MPI non-blocking point-to-point communication, and this happens for the data that do not belong in the same MPI-task (see inset in Figure \ref{fig:solvers_independence} - simulation snapshot). For the data belonging in the same memory space (same MPI-task), the framework skips any MPI-communication and retrieves them immediately from the local memory. The coupling of the two solvers and the subsequent communication introduces an order at which the various solvers should be executed. The execution steps of Palabos-npFEM can be found in Figure \ref{fig:solvers_steps}. Figure \ref{fig:stacked_percentages} summarises the average computational load per operation in Palabos-npFEM. The CPU-only version needs approximately $\times 3$ more CPU-cores to cover the GPU absence. A detailed performance analysis can be found in \cite{Kotsalos2020}.

\setlength{\parindent}{0pt}
\section*{Quality control}
%\textcolor{blue}{Detail the level of testing that has been carried out on the code (e.g. unit, functional, load etc.), and in which environments. If not already included in the software documentation, provide details of how a user could quickly understand if the software is working (e.g. providing examples of running the software with sample input and output data).}
Palabos-npFEM uses the default quality control tools integrated in Palabos. Palabos is hosted in GitLab, which provides continuous integration tools. From these tools, we have a test checking that the latest Palabos version compiles successfully. Currently, no unit testing framework is implemented. Nevertheless, we provide a number of example applications that any user can go through, and check if the various solvers are operational. Palabos-npFEM is extensively documented, and there are instructions for installing the software in any supported system, but also instructions for testing various example applications. See \textbf{Example Applications} sections below for more details on testing Palabos-npFEM library, and verifying its operational status. Furthermore, an extensive validation and verification of our framework has been performed in Kotsalos et al. \cite{Kotsalos2019,Kotsalos2020}.

\section*{Example Applications: Instructions}
Palabos-npFEM can be downloaded by cloning Palabos from the GitLab repository\footnote[1]{\url{https://gitlab.com/unigespc/palabos}}, and accessing the folder \texttt{examples} $\rightarrow$ \texttt{showCases} $\rightarrow$ \texttt{bloodFlowDefoBodies}. It contains the principal application (\texttt{bloodFlowDefoBodies.cpp}) of Palabos-npFEM. The principal application shows how to use Palabos \& npFEM in the FSI context, covering meticulously all the possible applications of the library. Instructions are provided to compile the software. The example applications presented below can be reproduced by using the different \texttt{xml} files provided in \texttt{bloodFlowDefoBodies} folder. For this, type the following in the command line after compiling the framework:
{\footnotesize
\begin{verbatim}
# CPU-only version
mpirun/mpiexec -n MPI_Tasks bloodFlowDefoBodies Example_Application.xml

# Hybrid CPU-GPU version
mpirun/mpiexec -n MPI_Tasks bloodFlowDefoBodies_gpu Example_Application.xml \
               NumberOfNodes NumberOfGPUsPerNode
\end{verbatim}
}
The locally provided \textbf{README.md} file helps the user run the applications in a step-by-step manner.

\setlength{\parindent}{15pt}
\section*{Example Application: Cell Packing}
\noindent To perform a cellular blood flow simulation, one needs to prepare the initial conditions, i.e. a flow field packed with blood cells (cell Packing). There exist various approaches, which usually require the use of an external tool to perform this initialisation. A standard solution is to place shrunken blood cells randomly in the flow field, and then let them grow back to their rest configuration. Another approach is to use tools that pack spheres or ellipsoids, and then replace the packed bodies with the blood cells (less accurate cell packing).

\begin{figure}[h]
    \centering
    \includegraphics[width=0.7\textwidth]{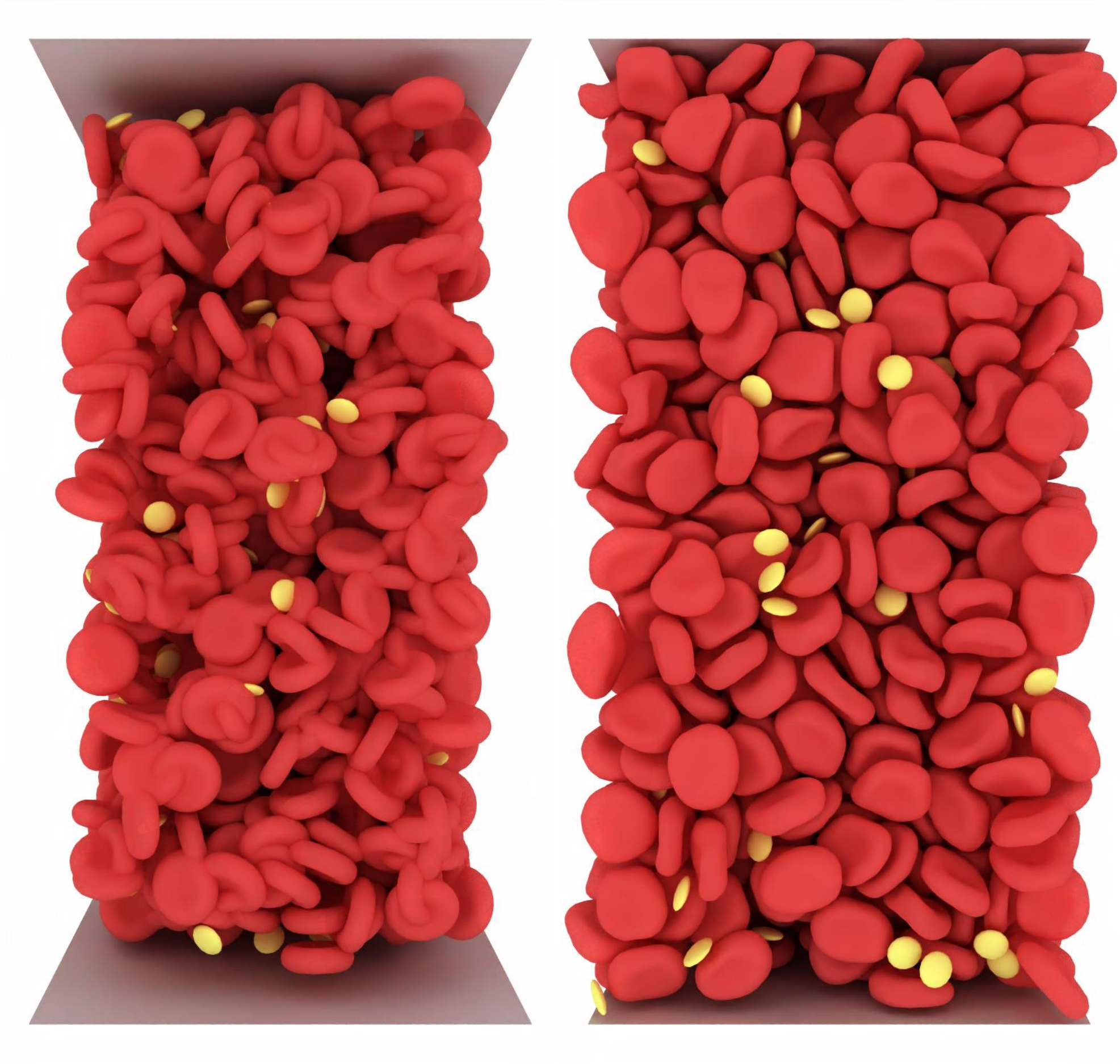}
    \caption{Cell packing as performed by Palabos-npFEM (case at 35\% hematocrit). The left inset shows the initial setup with unresolved interpenetrations, while the right one shows the resulted setup (resolved overlappings) after running the framework for a few thousand iterations.}
    \label{fig:cellPacking}
\end{figure}

Our approach falls closer to the standard solution, but instead of shrinking the blood cells, we randomly place them (random positions and orientations) at their rest configuration. Obviously, this leads to large overlappings/interpenetrations, which we anticipate to be resolved by the robust nature of the npFEM solver. In more details, npFEM is an implicit solver, which makes it capable of resolving very high deformations/interpenetrations with unconditional stability for arbitrary time steps. Taking advantage of this property, we run the framework for few thousand time steps, which indeed resolves the overlappings. Figure \ref{fig:cellPacking} presents the cell packing application as performed in a box at 35\% hematocrit. The cell packing application deactivates the branches of the code that deal with the fluid and the FSI (since there is no need for them), and instead uses the framework for efficiently detecting colliding neighbours and for executing the npFEM solver. The \texttt{cellPacking\_params.xml} (found in \texttt{bloodFlowDefoBodies} folder) provides options to perform the cell packing under various geometries and at different hematocrit. This novel approach introduces no further complexity to the framework, leading to a clean and efficient cell packing application (no need for external tool or additional code).

\setlength{\parindent}{0pt}
\section*{Example Application: Simulation of multiple blood cells}
\begin{figure}[h]
    \centering
    \includegraphics[width=\textwidth]{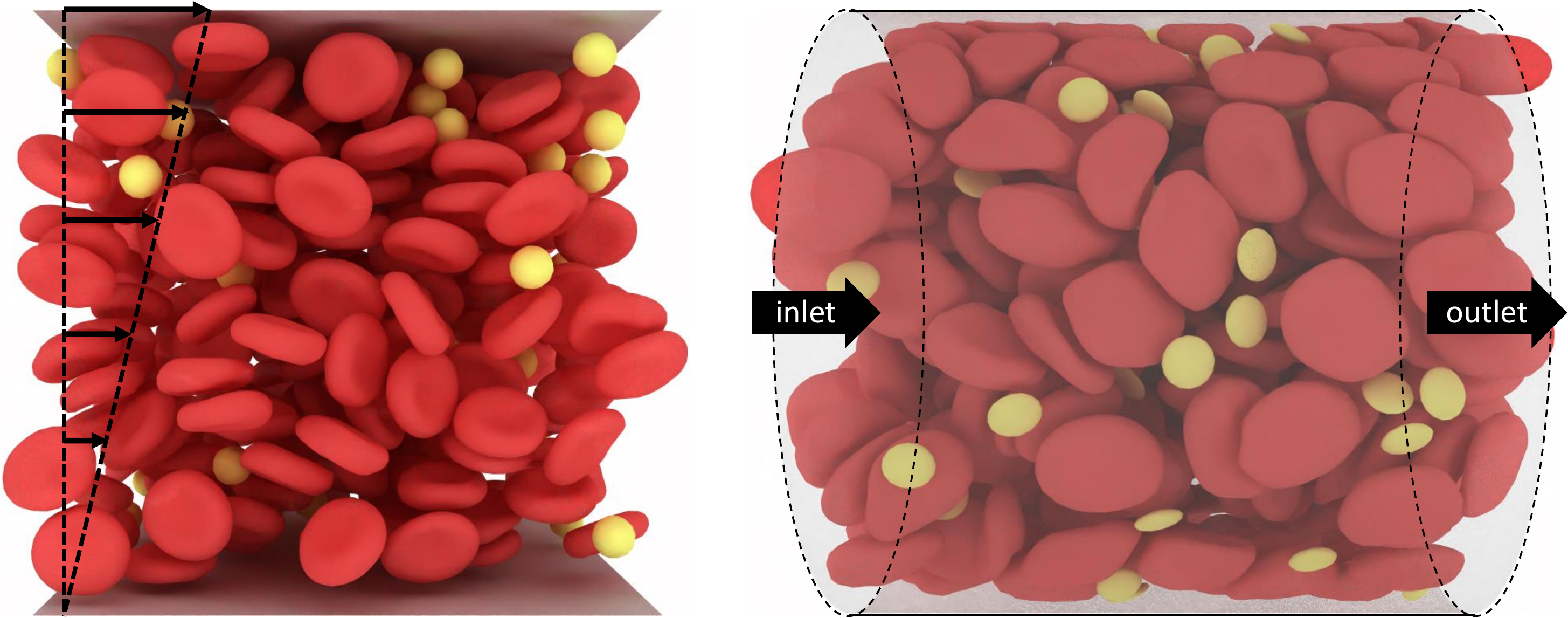}
    \caption{Simulation snapshots for a Couette (left, 20\% hematocrit) and a Poiseuille (right, 35\% hematocrit) flow. Through the provided \texttt{xml} files, one can choose the geometry, the hematocrit, and various other parameters.}
    \label{fig:simulations_box_tube}
\end{figure}

Having generated an initialised blood cell flow field, one can proceed to simulations of various flow regimes, e.g. tubular and shear flows. Figure \ref{fig:simulations_box_tube} presents two simulation snapshots as generated by running the \texttt{bloodFlowDefoBodies} application using the \texttt{shear\_params.xml, poiseuille\_params.xml} files (found in \texttt{bloodFlowDefoBodies} folder). It should be highlighted that the cell packing application is a prerequisite for running an actual simulation.

\section*{Example Application: RBC Collision at an obstacle}
\begin{figure}[h]
    \centering
    \includegraphics[width=0.4\textwidth]{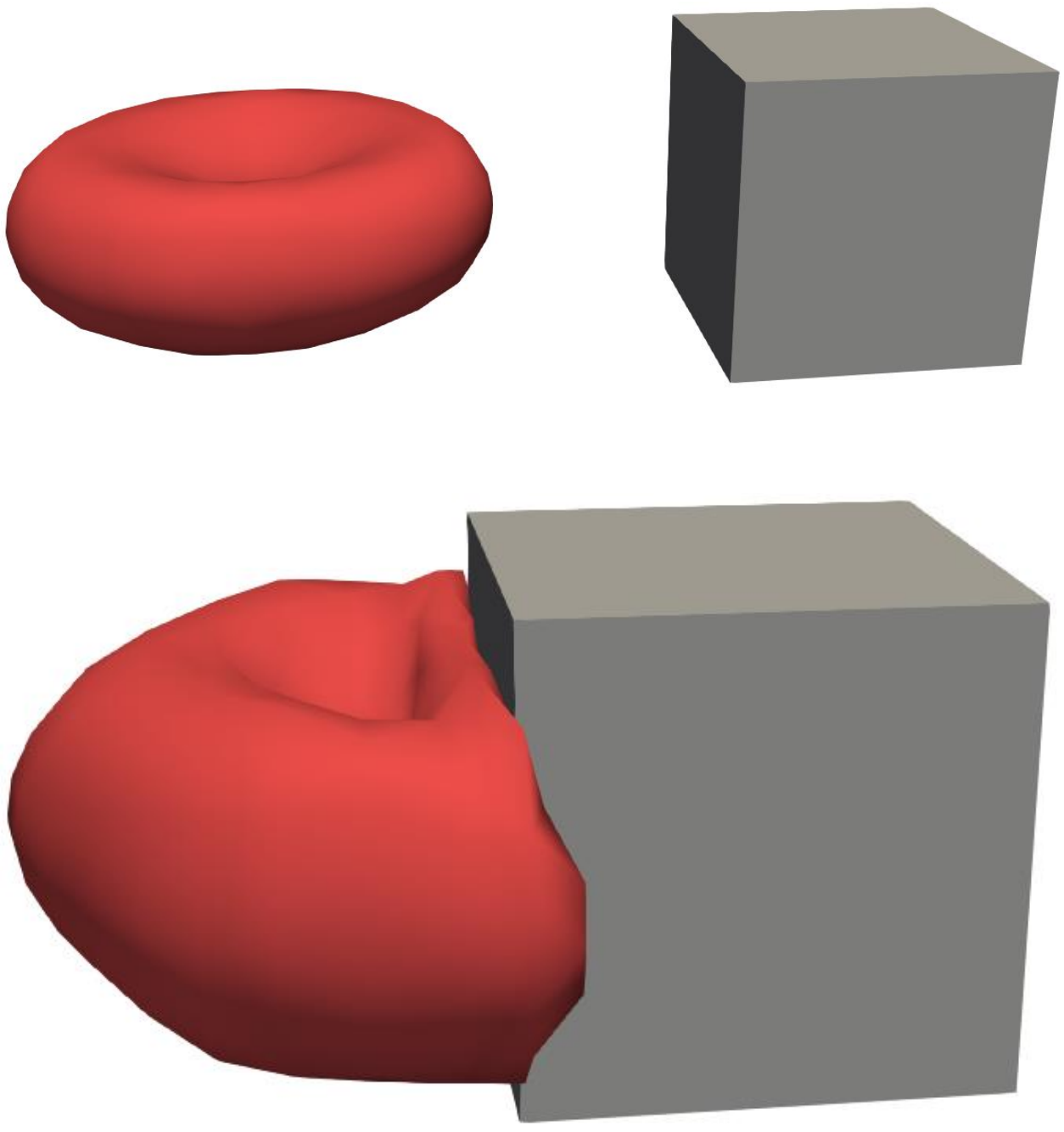}
    \caption{Simulation snapshot of a colliding RBC at an obstacle.}
    \label{fig:collision}
\end{figure}

This application (using the \texttt{obstacle\_params.xml}, found in \texttt{bloodFlowDefoBodies} folder) simulates a single RBC interacting with an obstacle. The position and orientation of the RBC can be tuned through an external file (see \texttt{initialPlacing.pos}). Various parameters can be tuned from the \texttt{xml} file, e.g. RBC viscoelasticity (\texttt{Calpha} parameter) and collision forces intensity (\texttt{collisions\_weight} parameter). This lightweight application provides a simple quality control for the robustness of the RBC material. Figure \ref{fig:collision} shows a snapshot from this application.

\setlength{\parindent}{15pt}
\section*{Rhino-Grasshopper environment for setting up new materials}
\noindent The framework/environment presented in this section is intended for \textbf{Windows only}. Setting up new materials, in a platform-agnostic way, is presented in the next section.

The npFEM library (derived from ShapeOp) can be used as a stand-alone library independently of Palabos. The library can be found by cloning Palabos from the GitLab repository\footnote[1]{\url{https://gitlab.com/unigespc/palabos}}. The source code is located in the folder \texttt{coupledSimulators} $\rightarrow$ \texttt{npFEM}. Inside this folder, one can navigate further to \texttt{npFEM\_StandAlone\_RhinoGH} folder, where we provide the option to compile npFEM as a stand-alone dynamic library. The produced \texttt{dll} can later be used in Rhino\footnote[2]{\url{https://www.rhino3d.com/}}-Grasshopper\footnote[8]{\url{https://www.grasshopper3d.com/}} environment. Rhino is a computer aided design software, and Grasshopper is a plug-in for Rhino intended for parametric design. Grasshopper uses the Python language for scripting various operations, and by using the Python standard foreign function library \texttt{ctypes}, one can call npFEM from within Rhino-Grasshopper. The locally provided \textbf{README.md} file helps the user setup the framework in a step-by-step manner.

The ShapeOp library provides a complete description in its documentation\footnote[3]{\url{https://www.shapeop.org/documentation.php}} on how to setup the Rhino-Grasshopper framework, but instead of ShapeOp, we will be using the compiled npFEM dynamic library. For legacy reasons with ShapeOp, we are using Rhino version 5, but the same principles apply to newer versions.

\begin{figure}[h]
    \centering
    \includegraphics[width=0.95\textwidth]{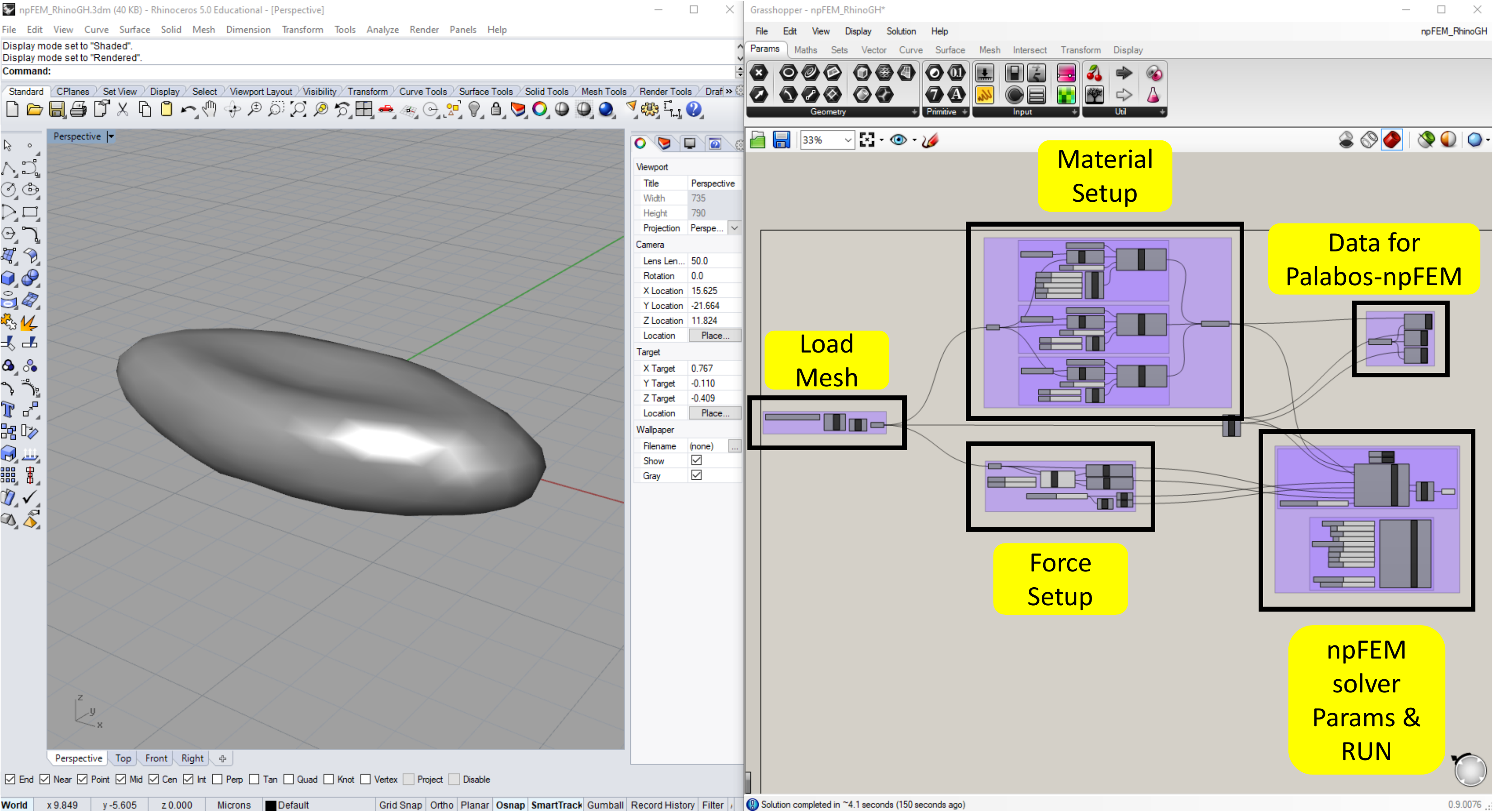}
    \caption{Rhino-Grasshopper environment calling the npFEM stand-alone dynamic library.}
    \label{fig:Rhino_GH}
\end{figure}

\begin{figure}[h]
    \centering
    \includegraphics[width=\textwidth]{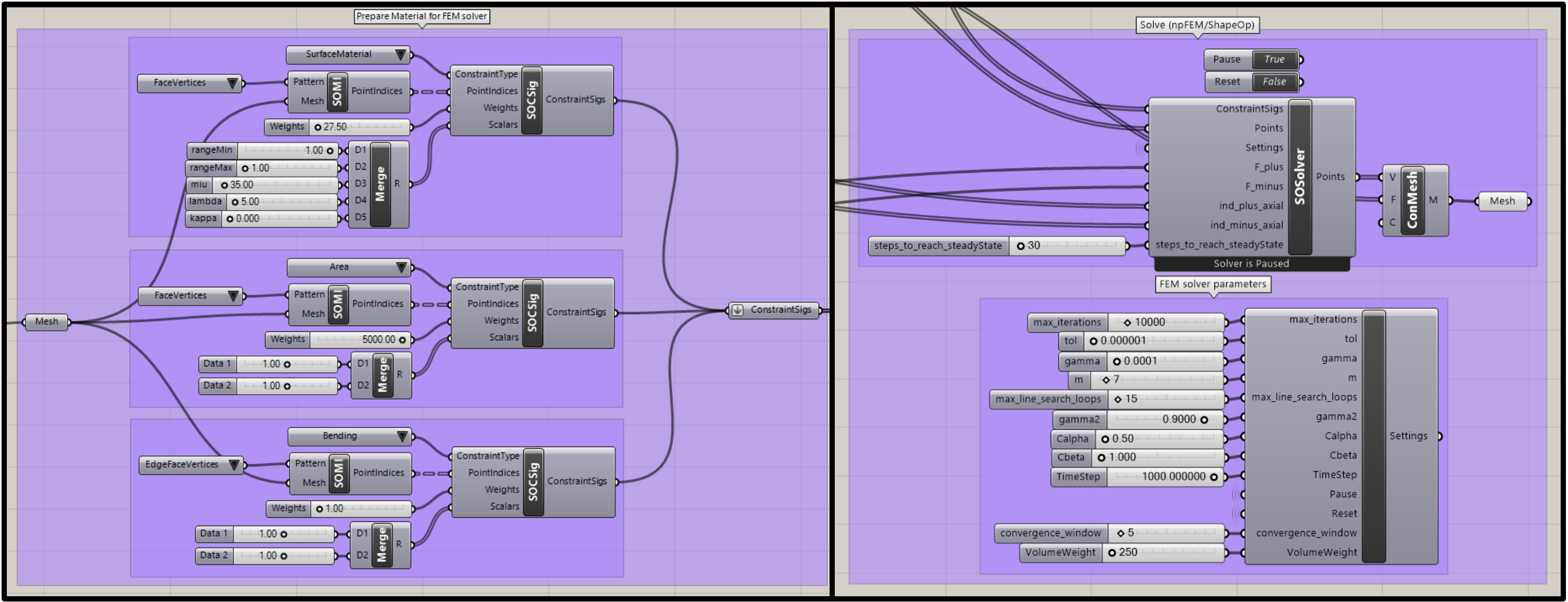}
    \caption{Components in the Grasshopper environment, from which one can modify the body's material, and run the npFEM solver.}
    \label{fig:GH_components}
\end{figure}

After setting up the environment, one can open the Rhino \& Grasshopper files provided in \texttt{npFEM\_StandAlone\_RhinoGH} $\rightarrow$ \texttt{npFEM\_RhinoGH}. Figure \ref{fig:Rhino_GH} presents the environment with a setup that shows a stretched RBC in the left pane (Rhino window), and in the right pane there is the Grasshopper window. The user can load various meshes, setup the material properties, add forces, and eventually run the npFEM solver. Figure \ref{fig:GH_components} presents a closer look to some critical Grasshopper components. The generation of RBC/PLT meshes can be done either in Rhino-Grasshopper (and any other CAD software), or through scripts using formulas to generate the investigated surfaces \cite{Boudjeltia2020}.

This environment offers an easy way to test and familiarise the user with the npFEM library. The users can experiment with different materials and solver parameters, and observe their impact on the deformed shape of a RBC. Additionally, one can modify the npFEM solver, and graphically observe if the modifications work as expected or not.

\setlength{\parindent}{15pt}
\section*{Setting up new materials: multi-platform}
\noindent Setting up new materials, in a platform-agnostic way, is more tedious than the previous solution. The Rhino-Grasshopper environment is the preferred way. The idea is to modify the file that encodes the material properties, which is located in \texttt{examples} $\rightarrow$ \texttt{showCases} $\rightarrow$ \texttt{bloodFlowDefoBodies} $\rightarrow$ \texttt{Case\_Studies}. In the current version of the library we provide template files for RBCs and PLTs discretised with 258 and 66 surface vertices, respectively. Extending to other bodies is a straightforward process (possible automation with minor scripting could be an option). The material is encoded in the \texttt{constraints.csv} file (found in \texttt{Case\_Studies} $\rightarrow$ \texttt{RBC/PLT}). This file encodes the various potential energies per finite element/triangle (per row). Summing the contributions of the elemental potential energies, one gets the potential energy of the whole body, which describes the body's response to deformations. Every elemental potential energy (a row in \texttt{constraints.csv}) has a weight and various parameters. Therefore, by tuning these parameters one can make the material stiffer or softer.

The tuning of all the other parameters is done through the provided \texttt{xml} files. Parameters of interest could be the ones responsible for the viscoelastic behaviour of the bodies (\texttt{Calpha, Cbeta} in the files), and the fluid characteristics (viscosity, density). A detailed explanation of the mechanics can be found in \cite{Kotsalos2019}.

\setlength{\parindent}{0pt}
\section*{(2) Availability}
\vspace{0.5cm}
\section*{Operating system}
Modern Linux \& Windows

\section*{Programming language}
C\texttt{++}11 and above

\section*{Additional system requirements}
Memory and disk space dependent on usage case.

\section*{Dependencies}
A complete list of prerequisites can be found in Palabos GitLab repository.\\
Linux:
\begin{itemize}
    \item A modern C\texttt{++} compiler
    \item make
    \item CMake ($\geq 3.0$)
    \item MPI (Message Passing Interface)
\end{itemize}
Windows:
\begin{itemize}
    \item Microsoft Visual Studio Community ($\geq 2015$)
    \item Microsoft C\texttt{++} compiler, installed with Visual Studio Community
    \item CMake ($\geq 3.0$)
    \item Microsoft MPI
\end{itemize}
The GPU branch of the npFEM solver supports NVIDIA GPUs with compute capability $\geq 6.0$ (extensively tested and validated). However, the code could support GPUs with compute capability $<6$ (tested but not fully validated), by replacing the atomic operations (e.g. \texttt{atomicAdd}) with ones based on \texttt{atomicCAS} (Compare and Swap). For more information, one should consult the CUDA toolkit documentation. In this case, the user should modify as well the CMake file to target the correct GPU architecture (currently set to \texttt{-arch=sm\_60}).

The periodic boundary conditions introduce a dependency on parallelisation. In more details, the directions with periodicity should be subdivided by at least two sub-domains (each sub-domain belongs to a different MPI task). This is due to how the area per surface vertex is computed, i.e. if there is no domain subdivision, then the crossing bodies (from the outlet to inlet) are considered as stretched (erroneous deformation, leading to code crash). On the contrary, when the outlet and inlet belong to different sub-domains (MPI-wise), then the crossing bodies are duplicated in memory and thus they are not considered as stretched/deformed. This means that Palabos-npFEM depends strictly on MPI, which is not a hard constraint given the computational intensity of blood flow simulations. Of course, this dependency could be eliminated by modifying/extending the library (area computation part).

\clearpage
\section*{List of contributors}
In addition to the paper authors, we wish in particular to acknowledge the contribution from the following person:
\begin{itemize}
    \item Joel Beny (University of Geneva), for the development of a major part of the GPU implementation of the npFEM solver.
\end{itemize}

\section*{Software location}

{\bf Archive}
\begin{description}[noitemsep,topsep=0pt]
	\item[Name:] Palabos-npFEM
	\item[Persistent identifier:] \url{https://doi.org/10.5281/zenodo.3965928}
	\item[Licence:] Palabos $\rightarrow$ AGPL v3 \& npFEM $\rightarrow$ MPL v2
	\item[Publisher:]  Christos Kotsalos
	\item[Version published:] 2.2.0
	\item[Date published:] 03/07/2020
\end{description}

{\bf Code repository}
\begin{description}[noitemsep,topsep=0pt]
	\item[Name:] Palabos
	\item[Persistent identifier:] \url{https://gitlab.com/unigespc/palabos.git}
	\item[Licence:] Palabos $\rightarrow$ AGPL v3 \& npFEM $\rightarrow$ MPL v2
	\item[Date published:] 03/07/2020 (v2.2.0)
\end{description}

\section*{Language}
English

\setlength{\parindent}{15pt}
\section*{(3) Reuse potential}
%\textcolor{blue}{Please describe in as much detail as possible the ways in which the software could be reused by other researchers both within and outside of your field. This should include the use cases for the software, and also details of how the software might be modified or extended (including how contributors should contact you) if appropriate. Also you must include details of what support mechanisms are in place for this software (even if there is no support).}
\noindent The Palabos-npFEM library gives special attention to modularity and low complexity. In more details, the software is designed based on a plug-and-play approach, where it is up to the user's preference to choose the individual solvers for the resolution of the various phases of blood. The Computational Biomedicine community is a vibrant and dynamic community, with numerous research contributions in various directions, thus we expect other researchers to possibly plug their own solvers in our platform and experiment with it. Starting point for extending and reusing Palabos-npFEM is the principal application located in \texttt{examples} $\rightarrow$ \texttt{showCases} $\rightarrow$ \texttt{bloodFlowDefoBodies} (utilised for the example applications). The end user can either deploy the library as is by executing this provided application, or build on top of it further functionalities/alterations.

Currently, the principal application that we provide treats simple geometries, i.e. box and tubular flows. However, its extension to more complicated geometries is well supported both by Palabos and npFEM. We consider that promising application areas for our software are provided by microfluidic devices and lob-on-a-chip systems, for which a large interest can be observed in the community.

Our library uses the CMake\footnote[4]{\url{https://cmake.org/}} tool for building and compiling. CMake is an open-source and cross-platform tool, which allows the libraries using it to be compiled in any supported platform. Thus, the users can deploy Palabos-npFEM in cross-platform environments (from personal computers/workstations to supercomputers) and speedup their development \& research.

Cellular blood flow simulations are extremely computationally expensive. For example \cite{Kotsalos2020}, to simulate a box of dimensions $50^3~\mu m^3$ under a shear flow at 35\% hematocrit, for physical time of 1 $s$, we need about 5 days in a high-end supercomputer (using 5 compute nodes, i.e. 12 cores and 1 GPU per node). However, an allocation of 5 consecutive days is rarely available in supercomputing centres. For this reason, we have developed an efficient check-pointing system, which allows the user to pause at any time the simulation, and restart seamlessly from where it previously stopped. This feature offers an attractive advantage for other researchers to actively use our library.

The library is specialised on cellular blood flow simulations, but its methodology could easily be applied to the simulation of other complex suspensions, and fluid-structure/solid interaction applications in general. A recent example is the simulation of Paragliders \cite{Lolies2019}, where the researchers used Palabos and a structural solver similar to npFEM. Thus we strongly believe that our library could be used as a building component for other research topics.

The Palabos library has a large and active community. Integrating npFEM into Palabos serves the purpose of sharing and exposing all the details with this global and dynamic group of researchers and engineers. The users can find support in the Palabos forum\footnote[1]{\url{https://palabos-forum.unige.ch/}}, and thus our library benefits from the same high-quality support mechanism that is already in-place for Palabos.

\setlength{\parindent}{0pt}
\section*{Acknowledgements}
We acknowledge support from the Swiss National Supercomputing Centre (CSCS, Piz-Daint supercomputer), the National Supercomputing Centre in the Netherlands (Surfsara, Cartesius supercomputer), and the HPC Facilities of the University of Geneva (Baobab cluster).

\section*{Funding statement}
This project has received funding from the European Union’s Horizon 2020 research and innovation programme under grant agreement No 823712 (CompBioMed2 project), and by the Swiss PASC project ``Virtual Physiological Blood: an HPC framework for blood flow simulations in vasculature and in medical devices''.

\section*{Competing interests}
The authors declare that they have no competing interests.

\clearpage
\printbibliography

\end{document}